\documentclass[journal=jacsat,manuscript=article]{achemso}

\usepackage{graphicx}%
\usepackage[T1]{fontenc}
\usepackage{upgreek}
\usepackage{braket}
\usepackage{subcaption}
\usepackage{amsmath}
\usepackage{amssymb}
\usepackage{hyperref}

\title{Excitonic valley effects in monolayer WS$_2$ under high magnetic fields}

\keywords{transition metal dichalcogenides, WS$_2$, 2D materials, excitons, valley splitting, valley polarization}

\author{Gerd Plechinger}
\email{gerd.plechinger@physik.uni-r.de}
\phone{++49 / (0)941 / 943 -2053}
\fax{++49 / (0)941 / 943 -4226}
\author{Philipp Nagler}
\affiliation{Institut f\"ur Experimentelle und Angewandte Physik,
	Universit\"at Regensburg, D-93040 Regensburg, Germany}
\author{Ashish Arora}
\affiliation{Institute of Physics and Center for Nanotechnology, University of M\"unster, 48149 M\"unster, Germany}
\author{Andr\'es Granados del \'Aguila}
\affiliation{High Field Magnet Laboratory (HFML-EMFL), Radboud University, Toernooiveld 7, 6525 ED Nijmegen, The Netherlands}
\alsoaffiliation{present address: Division of Physics and Applied Physics, School of Physical and Mathematical Sciences, Nanyang Technological University, Singapore 637371, Singapore}
\author{Mariana V. Ballottin}
\affiliation{High Field Magnet Laboratory (HFML-EMFL), Radboud University, Toernooiveld 7, 6525 ED Nijmegen, The Netherlands}
\author{Tobias Frank}
\affiliation{Institut f\"ur Theoretische Physik,
	Universit\"at Regensburg, D-93040 Regensburg, Germany}
\author{Philipp Steinleitner}
\affiliation{Institut f\"ur Experimentelle und Angewandte Physik,
	Universit\"at Regensburg, D-93040 Regensburg, Germany}
\author{Martin Gmitra}
\affiliation{Institut f\"ur Theoretische Physik,
	Universit\"at Regensburg, D-93040 Regensburg, Germany}
\author{Jaroslav Fabian}
\affiliation{Institut f\"ur Theoretische Physik,
	Universit\"at Regensburg, D-93040 Regensburg, Germany}
\author{Peter C. M. Christianen}
\affiliation{High Field Magnet Laboratory (HFML-EMFL), Radboud University, Toernooiveld 7, 6525 ED Nijmegen, The Netherlands}
\author{Rudolf Bratschitsch}
\affiliation{Institute of Physics and Center for Nanotechnology, University of M\"unster, 48149 M\"unster, Germany}
\author{Christian Sch\"uller}
\author{Tobias Korn}
\affiliation{Institut f\"ur Experimentelle und Angewandte Physik,
	Universit\"at Regensburg, D-93040 Regensburg, Germany}

\makeatletter
\let\acs@address@list\relax
\makeatother

\begin{document}

This document is the unedited Author's version of a Submitted Work that was subsequently accepted for publication in Nano Letters, copyright \textcopyright \, American Chemical Society after peer review. To access the final edited and published work see \\ \url{http://pubs.acs.org/articlesonrequest/AOR-tV2YRvMpYYy4x9b6jH3P}

\begin{abstract}
	Transition-metal dichalcogenides can be easily produced as atomically thin sheets, exhibiting the possibility to optically polarize and read out the valley pseudospin of extremely stable excitonic quasiparticles present in these 2D semiconductors. Here, we investigate a monolayer of tungsten disulphide in high magnetic fields up to 30\,T via photoluminescence spectroscopy at low temperatures. The valley degeneracy is lifted for all optical features, particularly for excitons, singlet and triplet trions, for which we determine the g factor separately. While the observation of a diamagnetic shift of the exciton and trion resonances gives us insight into the real-space extension of these quasiparticles, magnetic field induced valley polarization effects shed light onto the exciton and trion dispersion relations in reciprocal space. The field dependence of the trion valley polarizations is in line with the predicted trion splitting into singlet and triplet configurations.\\
	\textbf{Keywords:} transition metal dichalcogenides, WS$_2$, 2D materials, excitons, valley splitting, valley polarization \\
\end{abstract}
\maketitle


Monolayer transition-metal dichalcogenides (TMDCs) have had a tremendous impact on semiconductor physics in the last years. Due to their direct bandgap and a large interband dipole moment, atomically thin layers emit strong photoluminescence and represent a semiconducting supplement to the two-dimensional and zero-gap material graphene \cite{Splendiani2010, Mak2010}. Additionally, the dielectric confinement enables the formation and high stability of strongly bound excitons, charged excitons (trions) \cite{Mak2013a, Ross2013a} and even excitonic molecules \cite{You2015a, Plechinger2015a}. The conduction- and valence-band edges are located at the K-points in the Brillouin zone, where valley and spin degree of freedom are directly coupled \cite{Xiao2012}. This effect manifests itself in a circular dichroism \cite{Zeng2012, Mak2012}, i.e., excitons can be created in a specific valley by choosing the right excitation helicity, giving access to an additional internal degree of freedom. However, excitonic states with different valley indices are energetically degenerate, linked by time-reversal symmetry. This degeneracy can be broken by intense optical excitation via the optical Stark effect \cite{Sie2014} or by magnetic fields \cite{MacNeill2015, Li2014, Wang2015b, Aivazian2014, Srivastava2014, Mitioglu2015, Arora2016}. A direct control of the valley pseudospin energy is of outstanding importance for future quantum electronics applications based on the manipulation of the valley pseudospin \cite{Mak2014}. In recent experiments, the magneto-optical properties of single-layer diselenides like MoSe$_2$ \cite{MacNeill2015, Li2014, Wang2015b} and WSe$_2$ \cite{Aivazian2014, Srivastava2014, Mitioglu2015} have been investigated, including a lifting of the valley degeneracy and the extraction of exciton and trion g factors. Recent reports on high-field magneto-optics of transition metal disulfide monolayers are limited so far either to the field-induced rotation of the linear polarization of excitonic photoluminescence \cite{Schmidt2016} or to the investigation of CVD-grown materials exhibiting large linewidths \cite{Stier2016, Stier2016a, Mitioglu2016}. This precludes a separate analysis of the behavior of different excitonic quasiparticles like charge-neutral excitons and trions in magnetic fields, which is substantial for understanding the underlying mechanisms for magnetic-field-induced changes in the optical spectrum.

Here, we present low-temperature photoluminescence (PL) measurements on high-quality mechanically exfoliated monolayer WS$_2$ exhibiting extraordinarily sharp resonances originating from several distinct excitonic quasiparticles and bring these samples in high magnetic fields up to 30\,T. Charge-neutral excitons, singlet and triplet trions, as well as presumably phonon-related excitonic features emerging at large external fields are identified in the PL spectrum. We determine g factors of $\approx -4.3$ for exciton and triplet trion, and two regimes (low and high magnetic field) with different g factors for singlet trions. A diamagnetic shift for the excitons (singlet trions) on the order of 1\,meV (2\,meV) at 30\,T is observed. The extracted exciton radius of 25\,\AA$\:$is in excellent agreement with a theoretical model. Furthermore, we demonstrate the generation of valley polarization via a magnetic field. The complex interplay between different excitonic quasiparticles and the valley-orbit-splitted exciton and trion dispersions \cite{Yu2014, Glazov2015} turns out to be of major importance in the investigated sample and leads in the case of excitons to a predominant population of the energetically unfavorable valley. The singlet and triplet trion valley polarization can be explained with combined formation-rate-related and dispersion-related effects.

\begin{figure}
	\includegraphics*[width=0.7\linewidth]{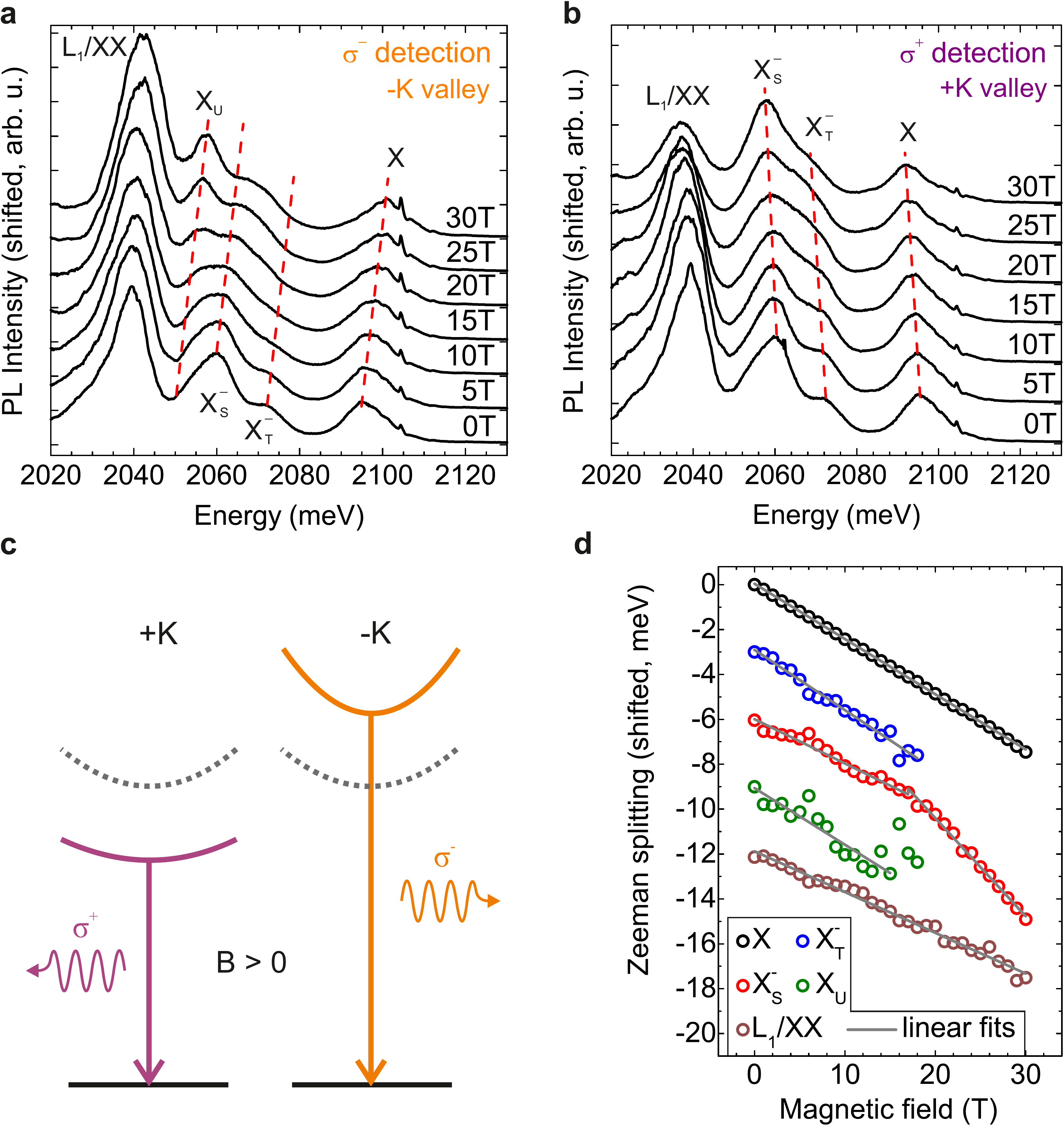}
	\caption{(a) $\sigma^-$-polarized photoluminescence spectra of single-layer WS$_2$ at $T=4.2$\,K for various magnetic fields up to 30\,T. (b) $\sigma^+$-polarized photoluminescence spectra of single-layer WS$_2$ at $T=4.2$\,K for various magnetic fields up to 30\,T. The red dashed lines in (a) and (b) serve as a guide to the eye for the energies of the X, X$^-_{T}$, X$^-_{S}$ and X$_{U}$ peaks. The narrow peaks around 2105\,meV are artefacts from the CCD and spectrometer. (c) Sketch of the +K and -K exciton dispersion under applied positive magnetic field. The grey dotted parabolas indicate the zero field dispersion. Due to exchange interaction, the dispersion steepness is different for +K and -K excitons in a magnetic field. (d) Zeeman splitting of the X (black), X$^-_{T}$ (blue), X$^-_{S}$ (red), X$_{U}$ (green) and L$_1$/XX peak (brown) as a function of the magnetic field. The data sets are shifted each by -3\,meV for better visibility. The grey lines indicate linear fits of the data points.}
	\label{fig:spectra}
\end{figure}

Figures \ref{fig:spectra}(a) and (b) highlight the photoluminescence spectra of a single-layer WS$_2$ flake (for a micrograph of the sample see Supplementary Fig. 1) at 4.2\,K and different magnetic fields from 0\,T to 30\,T applied in out-of-plane direction (Faraday geometry)  for left- ($\sigma^-$) (a) and right-circularly polarized ($\sigma^+$) detection (b). For all measurements presented in this work, the sample is excited via linearly polarized laser light at 2210\,meV, leading to an initially equal population of +K and -K valleys. At zero field, we can identify various peaks in the spectrum, details are published elsewhere \cite{Plechinger2015a, Plechinger2016a}. The charge-neutral exciton (X) appears at an energy of 2096\,meV, in good agreement with values in recent literature \cite{Plechinger2015a, Chernikov2014}. The negatively charged trion peak is split into an intravalley singlet (X$^-_{S}$) and an intervalley triplet trion peak (X$^-_{T}$) due to intervalley electron-hole exchange interaction \cite{Yu2014, Glazov2015, Jones2016, Plechinger2016a}. The energetic separation amounts to 10\,meV, with X$^-_{T}$ having the higher energy. 

Around 2040\,meV, a multiple peak structure (L$_1$/XX) stemming from defect-related excitons as well as biexcitons \cite{Plechinger2015a} is discernible in the spectra. However, the low excitation densities used in the experiment mainly preclude the observation of biexcitons. 

Applying a positive magnetic field leads to two major effects: first, all excitonic resonances shift in energy, depending on the detection helicity. Second, the relative intensities of the individual peaks vary significantly. The X$^-_{T}$ peak disappears at high magnetic fields in $\sigma^-$-polarized detection. At the same time, a new peak X$_{U}$ emerges 43\,meV below the X resonance. Theoretically, a biexciton binding energy in the same range as the trion binding energy has been predicted \cite{Kylanpaa2015, Zhang2015a}. However, only the L$_1$/XX feature exhibits a superlinear excitation-power-dependence which is consistent with biexcitonic PL (see Supplementary Fig. 2 and Supplementary Note 1). Therefore, a biexcitonic origin of the X$_{U}$ peak can be excluded. Instead, this feature can be tentatively interpreted as phonon-assisted transition of excitons, as the energetic difference between X and X$_{U}$ is in excellent agreement with the energy of the 2LA(M) Raman mode \cite{Berkdemir2013}. Nevertheless, more detailed studies are necessary in order to reveal the origin of this peak. Being energetically close to the trions and of low intensity, the X$_{U}$ peak is hardly visible in the zero field spectrum. In the $\sigma^+$-polarized spectra, it is barely discernible for all magnetic field strengths. The L$_1$/XX peak also dramatically decreases in intensity with increasing magnetic field in this detection polarization.

We can fit the exciton peak with a Lorentzian which accurately reproduces its lineshape. In $\sigma^+$ detection, we observe a redshift of the peak with increasing magnetic field, whereas in $\sigma^-$ detection, it experiences a blueshift. As we are probing the +K (-K) valley with $\sigma^+$ ($\sigma^-$)-polarized detection, we infer that the +K valley reduces its energy in a magnetic field whereas the -K valley increases in energy as illustrated in Figure \ref{fig:spectra}(c).

In single-layer TMDCs, three effects can lead to a magnetic-field-induced lifting of the valley degeneracy \cite{Li2014, Aivazian2014, Srivastava2014, Wang2015b}: spin Zeeman effect, valley Zeeman effect and the atomic orbital contribution (see Supplementary Fig. 3 and Supplementary Note 2). As bright excitons originate from conduction and valence band states having the same spin orientation, the spin Zeeman effect does not alter the energy difference of these bands resulting in unchanged exciton energies. The conduction band states do not carry an orbital magnetic moment, as they are comprised mainly of d orbitals with $m=0$. In contrast, the valence band maximum exhibits a valley-contrasting magnetic quantum number: $m=+2$ in the +K valley and $m=-2$ in the -K valley \cite{Xiao2012}. Hence, the +K valley exciton energy decreases by $2\mu_B B$, whereas the -K valley exciton energy increases by the same amount, with $\mu_B$ being the electron Bohr magneton and $B$ the magnetic flux density. This yields a valley splitting of $-4\mu_B B$, i.e., a g factor of $-4$. Defining the Zeeman splitting for excitons $E^X_{Zeeman}=E^X_{\sigma^+}-E^X_{\sigma^-}=g\mu_B B$, with the exciton energy in $\sigma^+$ ($\sigma^-$) detection $E^{X}_{\sigma^+}$ ($E^{X}_{\sigma^-}$), we extract an exciton g factor of $-4.25\pm0.05$ in our sample (see Fig. \ref{fig:spectra}(d), details are published elsewhere \cite{Schmidt2016}). This value differs slightly from 4, which would account for a pure atomic orbital contribution to the Zeeman splitting. The deviation could stem from the valley magnetic moment, which arises from the selfrotation of the Bloch wavepackets \cite{Chang1996, Xiao2007} and is oriented in opposite directions in the two valleys. In order to quantify the valley magnetic moment for excitons, we define $\Delta \alpha = \left( L^z_{c} - L^z_{v} \right) / \mu_B$, with the valley magnetic moments $L^z_{c}$ and $L^z_{v}$ of the conduction and valence bands. Including the atomic orbital contribution, we then have a net excitonic valley Zeeman splitting of $\Delta E^{X}_{Zeeman} = -2(2-\Delta\alpha)\mu_B B$ \cite{Aivazian2014}. In the framework of an \textit{ab initio}  DFT model, we have calculated $L^z_{c}$ and $L^z_{v}$ as well as the atomic orbital contributions which slightly deviate from the value of $- 2 \tau \mu_B B$, where $\tau$ is the valley index ($\pm1$ for $\pm$K), due to the admixture of p-type states to the band edges (see Supplementary Note 3 and Supplementary Fig. 4). It turns out that $\Delta \alpha$ has a negative sign and leads, together with a reduced orbital magnetic moment, to a g factor of -4.746. While the simplicity of the model precludes a thorough quantitative analysis, the calculations are in line with an absolute value of the g factor larger than 4, as observed in the experiment.

The trion peaks X$^-_{S}$ and X$^-_{T}$ in our spectra (Fig. \ref{fig:spectra}(a) and (b)) can be fitted by Gaussian functions. As X$^-_{T}$ is only visible as a shoulder and as its intensity decreases very fast with increasing magnetic field in $\sigma^-$-polarized detection, peak positions and intensities can be fitted only in the range from 0\,T to 18\,T. Defining the trion Zeeman splitting as the energy difference between the $\sigma^+$- and the $\sigma^-$-polarized PL peaks, we see a good agreement of the triplet trion g factor with the exciton g factor (Fig. \ref{fig:spectra}(d)). As the trions in mechanically exfoliated WS$_2$ monolayers are negatively charged excitons, we also have to consider the magnetic field-induced energy shifts for the conduction bands where the excess electrons reside. However, these energy levels are not directly visible in the PL spectrum, as the energy of the final state, which is the energy of the excess electron, has to be subtracted from the energy of the initial state in order to reproduce the experimentally observable peaks \cite{MacNeill2015}. The overall energy balance then reveals that the trion g factor should be identical to the exciton g factor. Surprisingly, the singlet trion g factor differs from the value of -4.3 and ranges between $-3.4\pm0.2$ from 0\,T to 18\,T and $-7.6\pm0.2$ from 18\,T to 30\,T. The origin of the dualistic behavior remains elusive so far. Large trion g factors exceeding the exciton g factor in absolute value could tentatively be explained by larger Berry curvature and hence also a larger valley magnetic moment for trions due to exchange interaction \cite{Srivastava2014}.

The X$_{U}$ feature exhibits a g factor of $-4.4\pm0.4$ (see Fig. \ref{fig:spectra}(d)), in good agreement with the exciton g factor. This makes an excitonic origin of the X$_{U}$ peak likely. The L$_1$/XX peak has a g factor of $-3.1\pm0.1$ (see Fig. \ref{fig:spectra}(d)), indicating a significant influence of defect-bound localized excitons.

\begin{figure}
	\includegraphics*[width=0.7\linewidth]{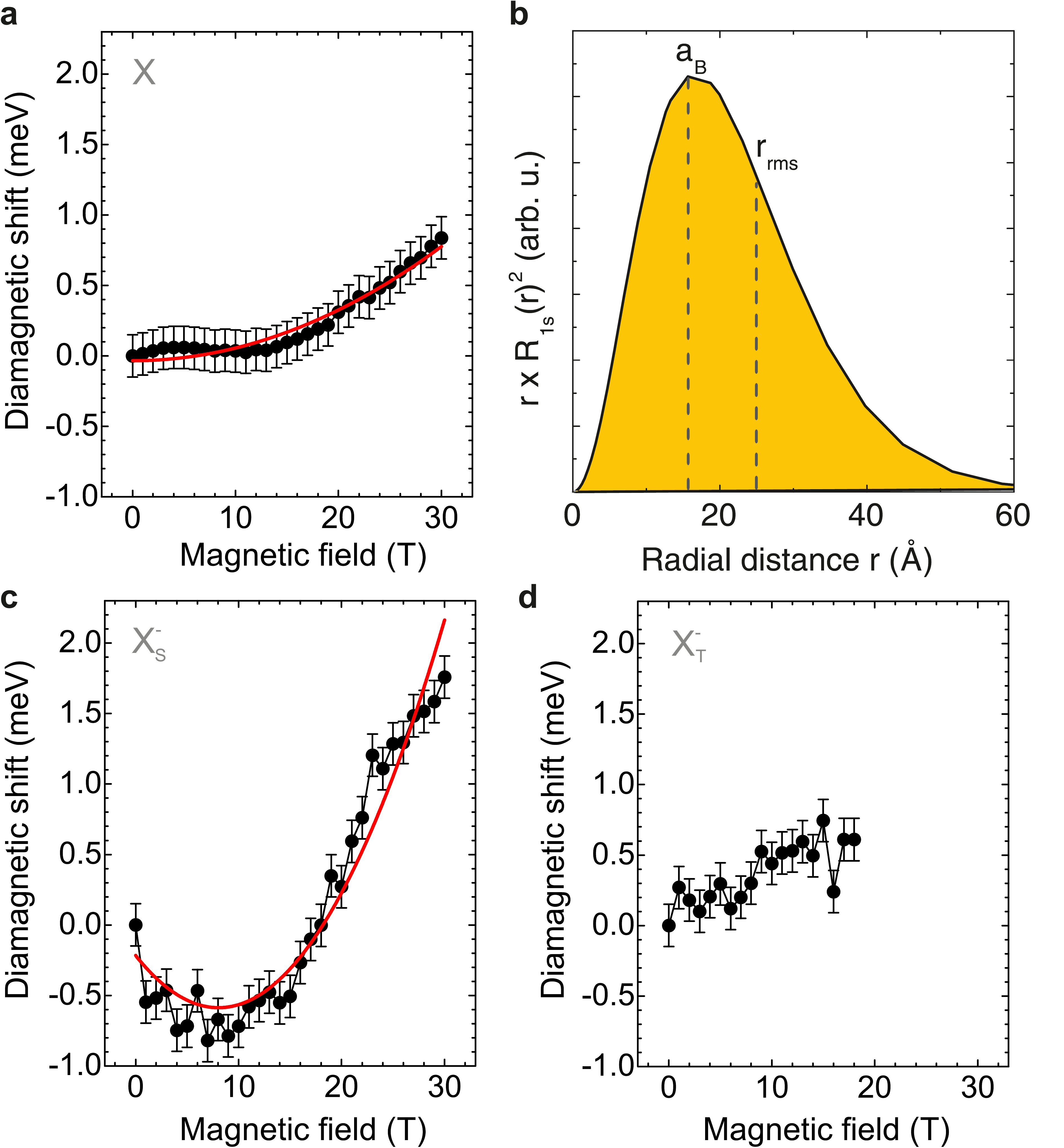}
	\caption{(a) Diamagnetic shift of the exciton resonance as a function of applied magnetic field. The red curve represents a parabolic fit of the data. (b) Calculated radial electron distribution for the 1s exciton ground state. $a_B$ is the Bohr radius, $r_{rms}$ the root-mean-square radius of the exciton. (c) Diamagnetic shift of the X$^-_{S}$ resonance as a function of applied magnetic field. The red curve represents the sum of a linear and a parabolic fit of the data. (d) Diamagnetic shift of the X$^-_{T}$ resonance as a function of applied magnetic field.}
	\label{fig:Dia}
\end{figure}

After having investigated the Zeeman splitting of 2D excitons, we now discuss the change of the exciton energy averaged over the two valley configurations. The Larmor term in the Hamiltonian of the system gives rise to a diamagnetic shift of an excitonic resonance, which manifests itself as a quadratic increase of the exciton energy in a magnetic field. Our good sample quality allows for the clear separation of the exciton spectral feature from other quasiparticle peaks related to trions or localized excitons. Therefore, we can investigate the diamagnetic shift stemming solely from charge-neutral excitons, in contrast to previous reports \cite{Stier2016, Stier2016a}, where it could be observed only for a spectrally broad ensemble of different excitonic quasiparticles. Fig. \ref{fig:Dia}(a) displays the energy shift of the exciton averaged over the two detection helicities as a function of $B$. The data can be fitted with a quadratic function following the equation $E^X_{dia} = aB^2$, with a factor $a=0.9\cdot10^{-6}\,$eVT$^{-2}$. This value is slightly larger than for a broad spectral feature with different excitonic portions, as reported by Stier et al. \cite{Stier2016, Stier2016a, Stier2016b} . The diamagnetic shift in units of eV can be described by the following equation \cite{Miura2008, MacNeill2015}:
\begin{equation}E^X_{dia}= \frac{e^2 r^2_{rms}}{8\mu} \cdot B^2 . 
\label{eq:dia}\end{equation}
Using a reduced effective mass of $\mu=0.15\,m_e$ \cite{Kormanyos2015} and the experimentally determined factor $a$, we get a root-mean-square exciton radius $r_{rms}$ of 25\,\AA. In order to compare this experi\-mental value with theory, we numerically solve a modified Schr\"odinger equation for 2D excitons \cite{Berkelbach2013, Chernikov2014, Poellmann2015} (see Supplementary Note 4). Hereby, we use a potential derived by Keldysh \cite{Keldysh1976} and take into account the additional dielectric screening by the substrate via an increased screening length \cite{Chernikov2014}. The solution for a charge-neutral exciton in a substrate-supported monolayer of WS$_2$ gives us a binding energy of 312\,meV, in line with other reports \cite{Chernikov2014, Chernikov2015a}. The calculated radial distance of the electron in the 1s ground state, $r \cdot R_{1s}(r)^2$, is highlighted in Fig. \ref{fig:Dia}(b). The maximum of the distribution appears at 16\,\AA $\:$ and corresponds to the Bohr radius $a_B$. The root-mean-square radius $r_{rms}$ is slightly larger and lies at 25\,\AA, in excellent agreement with our experimentally measured value. Hence, the spatial extension of the exciton can be directly extracted from our measurements and is strongly supported by a simple 2D exciton model. 

The X$^-_{S}$ peak also shows a diamagnetic shift, as plotted in Fig. \ref{fig:Dia}(c). Interestingly, we observe first a red shift up to 9\,T, followed by a more pronounced blue shift. Such a behavior for negatively charged trions has been reported already in conventional III-V-semiconductor systems \cite{Shields1995} and has been attributed to the combined effect of a magnetic-field-induced increase in binding energy, having a linear dependence on $B$ in first approximation, and of the diamagnetic shift, which has a quadratic dependence on $B$ and dominates at high fields \cite{Sandler1992}. The corresponding fit is shown as the red line in Fig. \ref{fig:Dia}(c). The prefactor $a$ for the diamagnetic shift is $5.7 \cdot 10^{-6}\,$eVT$^{-2}$ and consequently significantly larger than for the charge-neutral excitons. This finding hints to larger spatial extensions of the trions in comparison to excitons.

For the X$^-_{T}$ feature, a slight blue shift can be observed for the mean peak position (see Fig. \ref{fig:Dia}(d)). However, the data does not allow for a detailed analysis.

\begin{figure}
	\includegraphics*[width=0.7\linewidth]{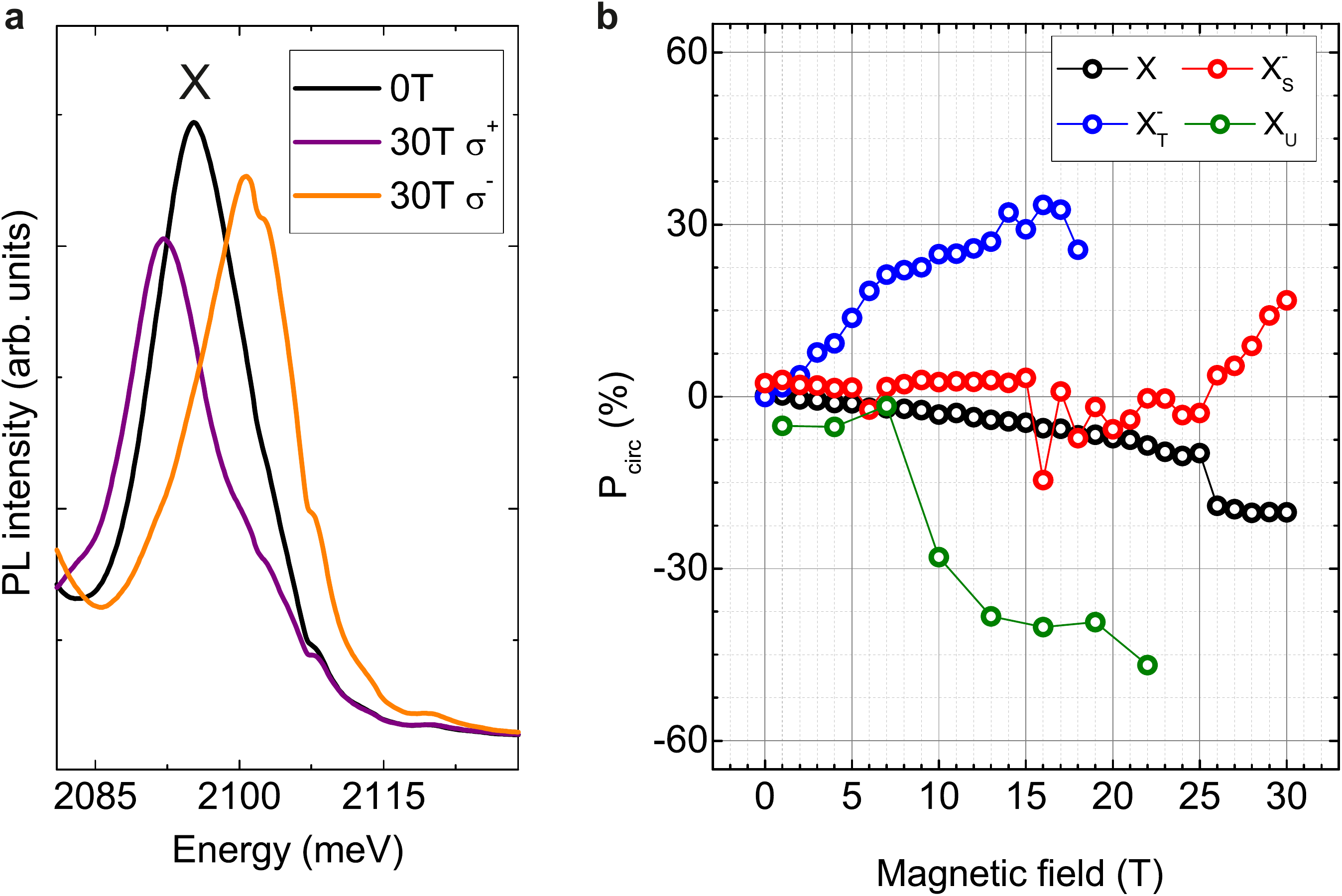}
	\caption{(a) Photoluminescence spectra of the exciton resonance without magnetic field (black curve), at $B=30\,$T for $\sigma^+$-polarized detection (purple curve) and at $B=30\,$T for $\sigma^-$-polarized detection (orange curve). (b) Magnetic field induced valley polarization degree $P_{circ}$ for X (black circles), X$^-_{T}$ (blue circles), X$^-_{S}$ (red circles) and X$_{U}$ (green circles) as a function of magnetic field. To increase the signal-to-noise ratio, the polarization degree of the X$_{U}$ peak has been averaged over 3 data points.}
	\label{fig:ValleyPol}
\end{figure}

\begin{figure}
	\includegraphics*[width=0.6\linewidth]{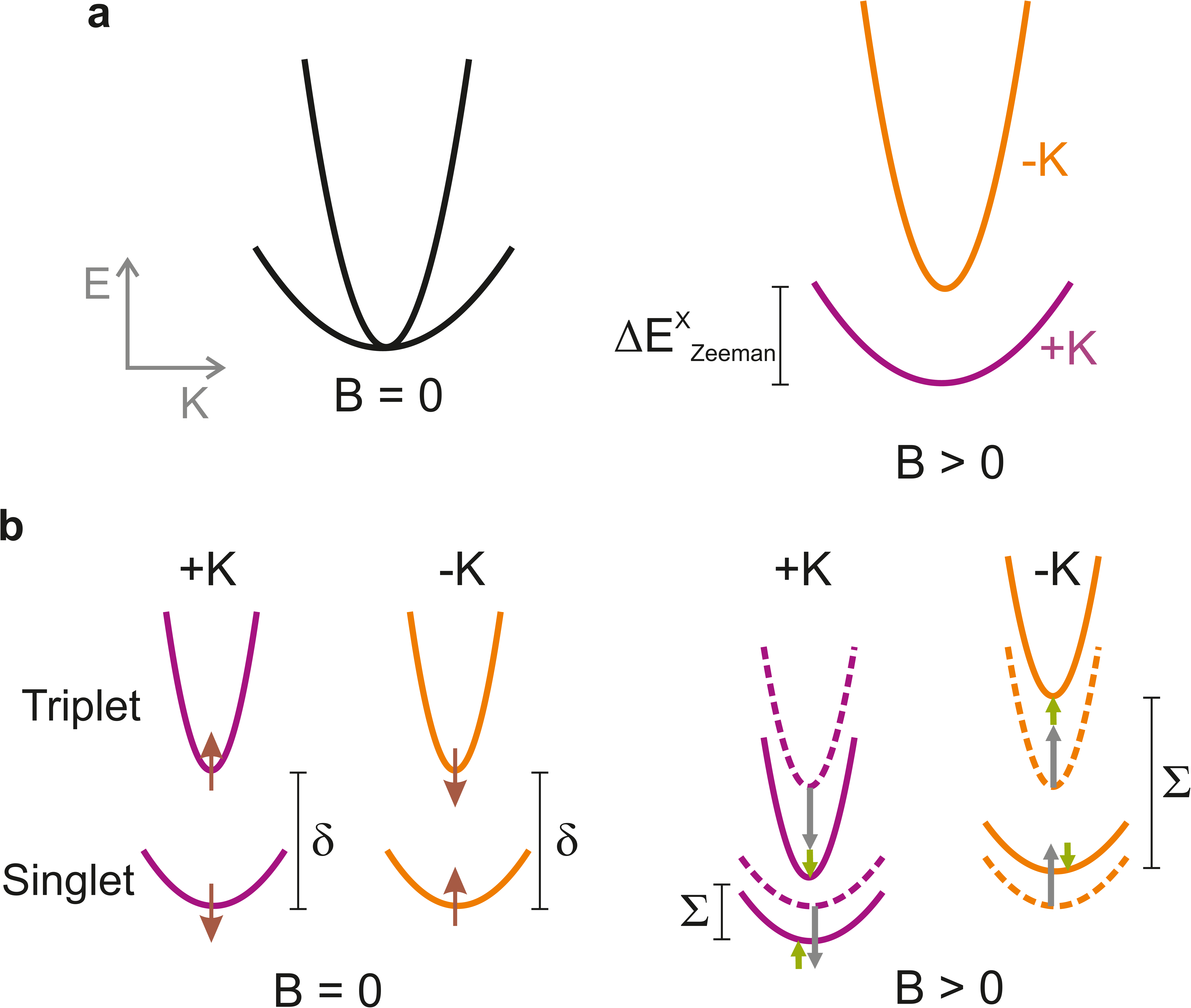}
	\caption{(a) Sketch of the exciton dispersion at zero magnetic field (left) and positive magnetic field (right). Black color indicates unpolarized states, orange -K valley polarized and purple +K valley polarized states. (b) Trion dispersion at zero (left) and finite positive magnetic field (right). The branches are split by the energy $\delta$ already without an external field. The trions are assigned to the valley, where the radiatively recombining electron-hole-pair is located. The spin orientation of the excess electron is indicated by the brown arrows. In a positive magnetic field, the trion energies shift according to the Zeeman shift of the electron-hole-pair (grey arrows) and the Zeeman shift of the excess electron conduction bands (green arrows). The energy difference between the two dispersion branches is quantified by the variable $\Sigma$. The dashed lines represent the zero-field dispersion.}
	\label{fig:ValleyPolScheme}
\end{figure}

When comparing the exciton peaks in PL spectra at finite magnetic fields for different detection helicities as shown in Fig. \ref{fig:ValleyPol}(a), one can see that, interestingly, the valley that shifts to higher energies gets polarized, and not the reverse, as reported for MoSe$_2$ \cite{MacNeill2015, Li2014, Wang2015b}. This striking behavior is counterintuitive, as one would expect that the valley with the lower energy is populated more. As pointed out in several references \cite{Aivazian2014, Yu2014, Glazov2015}, the exciton dispersion splits up into two branches which are energetically degenerate at $K=0$ and zero magnetic field, with the exciton momentum $K$ (see Fig. \ref{fig:ValleyPolScheme}(a)). The upper exciton branch has a steeper dispersion than the lower one, that means the bright exciton emission originates mainly from the upper branch, because for the excitons excited at higher energies a smaller change of their momentum is necessary to scatter into the light cone. In a magnetic field, the degeneracy of the two branches at $K=0$ is lifted and they get valley-polarized \cite{Aivazian2014} (see Fig. \ref{fig:ValleyPolScheme}(a)). This introduces an additional mechanism for bright exciton polarization, where the state with the lowest energy is expected to have a more efficient emission due to thermal population. It is reversed to the one related to dispersion steepness, discussed above, and represents, for example, the dominant mechanism in MoSe$_2$ \cite{Wang2015b}. In a simplified picture, the interplay between the two described processes, the first associated with the dispersion relation, the second with the lifted energy degeneracy of the two valleys, yields then the total magnetic field induced valley polarization. However, the observed polarization effects are the result of a dynamical equilibrium of excitons, different trion species and dark excitons \cite{Zhang2015b, Plechinger2016a}. Due to the magnetic-field-induced energy shifts, formation rates as well as relaxation rates might change separately for the individual quasiparticles \cite{Jeukens2002}. Also, the intervalley relaxation rates might exhibit a field dependence. Therefore, performing steady-state PL measurements that give only limited information on the dynamics, we refrain from prioritizing one of the above mentioned explanations for the anomalous exciton valley polarization. 

Magnetic-field-induced exciton valley polarization degrees $P_{circ} = \left(I_{\sigma^+}-I_{\sigma^-}) / (I_{\sigma +}+I_{\sigma -}\right)$, with $I_{\sigma \pm}$ being the integrated PL intensity of $\sigma^+$ ($\sigma^-$)-polarized light, of only -10$\%$ to -20$\%$ can be generated at a field of 30\,T. Note that the sign of $P_{circ}$ is negative, indicating a polarization of the energetically unfavorable valley. 

Fig. \ref{fig:ValleyPol}(b) summarizes the circular polarization degrees for all investigated spectral features. The discontinuity at 24\,T may be related to a slight change in the focus position or in the measurement position and does not affect the conclusions drawn in this section.

The X$_{U}$ peak shows a relatively large negative valley polarization up to $-50\,\%$ at the investigated field range (see Fig. \ref{fig:ValleyPol}(b)), having the same sign as for the charge-neutral exciton, which indicates, that for this peak similar polarization mechanisms as for the exciton are active.

In the case of singlet trions, we observe mainly a negligible magnetic-field-induced valley polarization, as depicted in Fig. \ref{fig:ValleyPol}(b). For triplet trions, $P_{circ}$ increases monotonically with the applied field, resulting in a polarization degree of $+30\%$ at around 15\,T. The positive sign of $P_{circ}$ indicates that the X$^-_{T}$ PL is emitted preferentially from the +K valley, which has the lowest energy. 

The degeneracy of the two dispersion branches at $K=0$, introduced above in the framework of charge-neutral excitons, is already lifted at zero magnetic field for the trions \cite{Yu2014, Aivazian2014}, with an energetic separation of $\delta=10$\,meV \cite{Plechinger2016a}. In order to investigate the shifts of the trion branches in a magnetic field, we have to take into account the valley index of the exciton and the energy shift of the excess electron conduction band, either in the same valley as the exciton (singlet trions), or in the opposite valley (triplet trions) \cite{Aivazian2014}. The corresponding scheme that arises from the theoretically supported assumption that the valley magnetic moment is larger than and reversed to the spin magnetic moment (see Supplementary Note 2) is drawn in Fig. \ref{fig:ValleyPolScheme}(b). In consequence, the spin-up conduction band state in the -K valley shifts to lower energies, whereas the opposite holds for the spin-down conduction band state in the +K valley (see Supplementary Fig. 3).
As the magnetic field is increased, the triplet branch in the +K valley reduces its energy, while the reverse holds for the -K valley. This means that the +K valley is energetically more favorable for triplet trions, which leads to a large valley polarization with opposite sign as compared to excitons. In order to explain the negligible valley polarization of the singlet trions, we have to look at the energetic difference $\Sigma$ between the two trion branches. In the +K valley, singlet and triplet branch approach each other, making the triplet branch energetically more favorable than without applied field. Given the smaller velocity distribution of the triplet trions, this leads to an increase of the triplet formation rate at the expense of the singlet formation rate. We have the reverse process in the -K valley, where the splitting $\Sigma$ increases with the magnetic field, yielding a larger singlet emission rate. This mechanism, which is driven by the magnitude of $\Sigma$, is working in the opposite direction than the polarization effect that stems from the pure energetic movement of the singlet branches in the K valleys. Hence, both mechanisms, assuming equal strength, cancel out giving rise to a zero polarization of the X$^-_{S}$ feature for almost all investigated magnetic fields.

The experiments presented here support the theoretical prediction of a splitting of the trion dispersion \cite{Yu2014, Glazov2015} and give new insights into 2D exciton physics in atomically thin transition metal dichalcogenides. Moreover, we have the possibility for tuning the singlet trion to triplet trion intensity ratio. Applying a magnetic field to a single-layer of WS$_2$ enables therefore additional control mechanisms for Coulomb-bound quasiparticle complexes in a two-dimensional semiconductor sheet.

In conclusion, we have investigated the photoluminescence of high-quality mechanically exfoliated monolayers of WS$_2$ in high magnetic fields. Owing to the narrow linewidth in our samples, we are able to observe a Zeeman splitting of +K and -K valley quasiparticles and to extract individually the g factors of charge-neutral excitons, intravalley singlet, intervalley triplet trions, a newly emerging and highly polarized feature 43\,meV below the exciton resonance, and of localized excitons. Furthermore, the diamagnetic shift of charge-neutral excitons, singlet and triplet trions can be analyzed separately. From these measurements, we deduce an exciton radius of 25\AA, which is in excellent agreement with the theoretical description of the excitonic quasi particle. The magnetic-field-induced valley polarization of excitons, singlet and triplet trions is in line with the picture of a longitudinal-transverse splitting of the exciton and trion dispersion. The polarization effects for the trions provide strong evidence for the correct assignment of the singlet and triplet trion feature. The results of this work related to the splitting of the energy degeneracy of the two distinct valleys as well as to the valley polarization effects in a magnetic field contribute to a more detailed understanding of the nature of the quasiparticles in these two-dimensional semiconductors, which is of outstanding importance for the realization of future valleytronic devices, where a complete control over the valley pseudospin is required.

\section*{Methods}
\subsection*{Sample preparation}
Single-layer flakes of WS$_2$ are first mechanically exfoliated from bulk crystals (hq Graphene inc.) onto PDMS substrates. Via a deterministic transfer process \cite{Castellanos2014}, the flakes are placed on the final substrate, a Si chip with 300\,nm SiO$_2$ capping layer and lithographically defined Au-markers on top. 
\subsection*{Magneto-PL spectroscopy}
The sample is mounted onto a x-y-z piezoelectric stage in a probe tube filled with He exchange gas. In the cryostat, it is cooled down to 4.2\,K. Magnetic fields up to 30\,T can be applied with a resistive magnet in Faraday geometry. For optical excitation, laser light at an energy of 2210\,meV is focussed onto the sample via a microscope objective in the probe tube to a spot size of $\approx 4$\,$\mu$m, resulting in an excitation density of 0.4\,kWcm$^{-2}$. The backscattered PL is guided to the spectrometer using a non-polarizing cube beam splitter. We use a liquid-nitrogen-cooled CCD chip to detect the dispersed light. The polarization of the PL is analyzed with a quarter-wave plate and a linear polarizer. 

\section*{Associated Content}
\subsection*{Supporting Information}
Micrograph of the investigated monolayer WS$_2$ sample, excitation-power dependent magneto-PL measurements, description of the magnetic-field-induced energy shifts of the lower-energy conduction bands, valley and orbital Zeeman angular momentum from ab-initio calculations, model for the theoretical description of a 2D exciton.

\section*{Author Information} 
\subsection*{Author contributions}
T.K., R.B., P.C.M.C, and G.P. conceived the experiments. G.P., P.N., A.A., A.G.A. and M.V.B. performed the experiments. T.F., M.G. and J.F. performed the DFT calculations, P.S. calculated the exciton model. G.P., C.S. and T.K. analyzed the data. G.P. wrote the manuscript with input from all authors.
\subsection*{Notes}
The authors declare no competing financial interests.

\section*{Acknowledgements}
The authors gratefully acknowledge financial support by the DFG via KO3612/1-1, GRK1570 and SFB689, the A. v. Humboldt foundation, and support of HFML-RU/FOM, member of the European Magnetic Field Laboratory (EMFL).


\newpage
\setcounter{figure}{0}
 \renewcommand{\figurename}{\footnotesize Supplementary Figure } 
	\begin{center}
		\large{Excitonic valley effects in monolayer WS$_2$ }
		\\ \large{under high magnetic fields} 
		\\ \large{- Supplementary Information -}
		\\ \small{Gerd Plechinger, Philipp Nagler, Ashish Arora, Andr\'es Granados del \'Aguila, Mariana V. Ballottin, Tobias Frank, Philipp Steinleitner, Martin Gmitra, Jaroslav Fabian, Peter C. M. Christianen, Rudolf Bratschitsch, Christian Sch\"uller, and Tobias Korn}
	\end{center}
	
	\section{Supplementary Figures}
	
	\begin{figure}[h]
		\includegraphics*[width=0.3\linewidth]{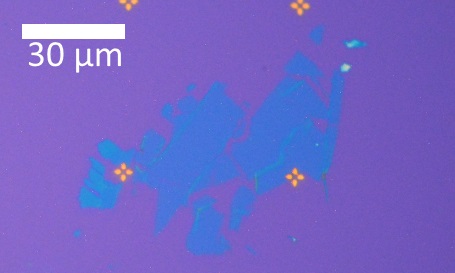}
		\caption{\footnotesize\textbf{Microscope image of the sample.} Micrograph of the investigated monolayer WS$_2$ sample on top of a Si/300\,nm thick SiO$_2$ substrate. The distance between two Au markers (crosses) is 50\,$\mu$m.}
		\label{fig:Supp_Mikroskopbild}
	\end{figure}
	\begin{figure}[h]
		\includegraphics*[width=0.4\linewidth]{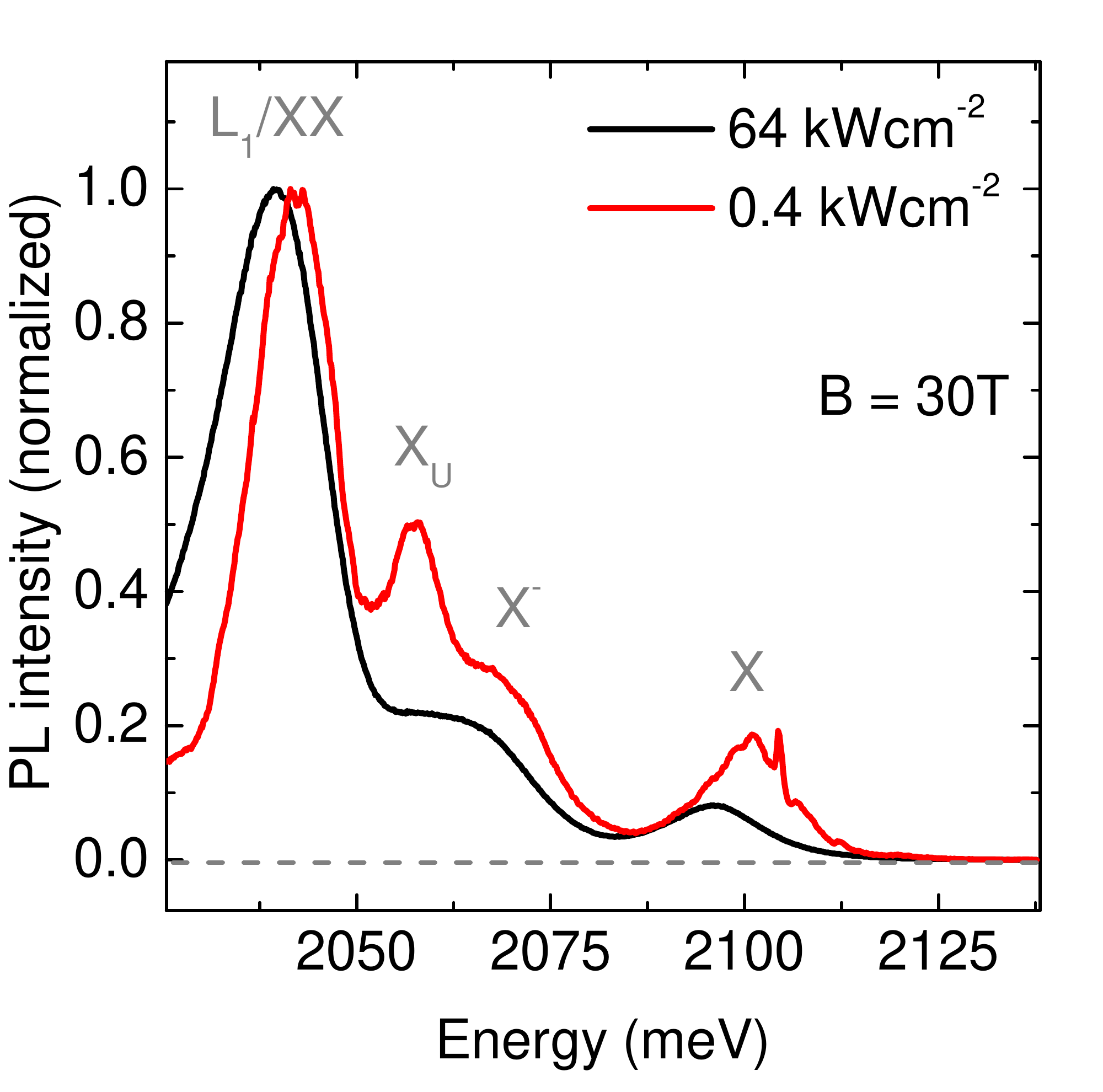}
		\caption{\footnotesize\textbf{Power-dependence of the PL at 30\,T.} Normalized PL spectra for excitation densities of 0.4\,kWcm$^{-2}$ (red curve) and 64\,kWcm$^{-2}$ (black curve) taken at a temperature of 4.2\,K and a magnetic field of 30\,T.}
		\label{fig:Supp_PowerComparison}
	\end{figure}
	\begin{figure}[h]
		\includegraphics*[width=0.8\textwidth]{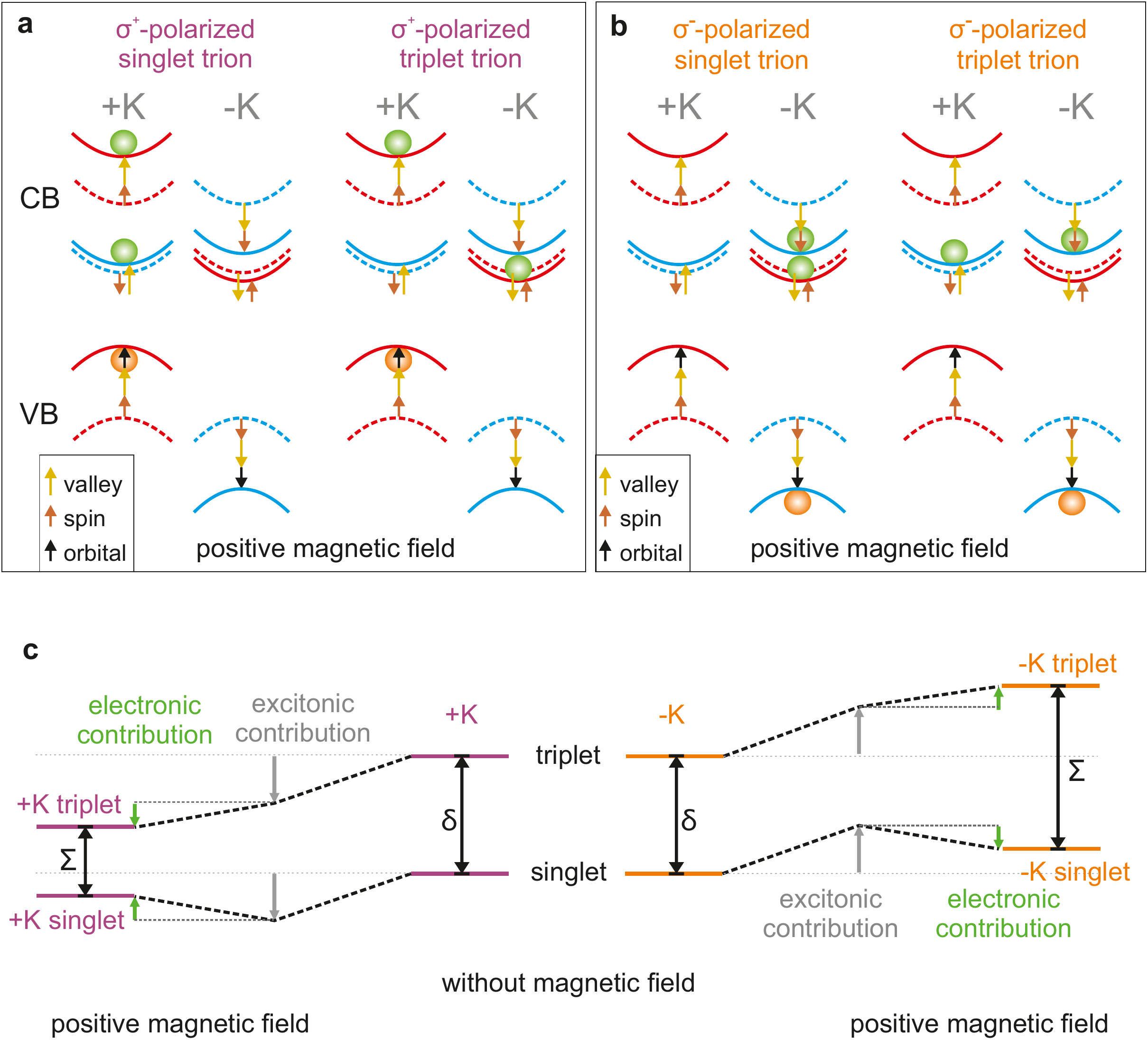}
		\caption{ \footnotesize \textbf{Scheme of the magnetic-field-induced shifts of the band and trion energies.} (a) Band diagram of monolayer WS$_2$ around +K and -K points under a positive out-of-plane magnetic field. The spin-up (spin-down) bands are shown in red (blue), the lower-energy spin-split valence band is omitted. The zero-field energies of the bands are denoted in dashed lines. The different contributions for the overall energy shifts are depicted as the brown, dark yellow and black arrows, corresponding to the spin magnetic moment, the valley magnetic moment and the atomic orbital magnetic moment, respectively. In the left (right) part of the subfigure, the electron configuration for the $\sigma^+$-polarized intravalley singlet-trion (intervalley triplet-trion) is displayed. Conduction-band electrons are symbolized in green, missing valence-band electrons in orange. (b) Electron configuration for the $\sigma^-$-polarized intravalley singlet-trion (left part of the subfigure) and intervalley triplet-trion (right part of the subfigure). (c) Energy-level diagram for the singlet and triplet trion energies in +K (purple) and -K valleys (orange) under a positive magnetic field. The shift due to the change in the exciton energy is indicated by the grey arrow, the shift due to the change in the lower conduction-band energy by the green arrow. The singlet-triplet splitting without (with) magnetic field is described by the variable $\delta$ ($\Sigma$).}
		\label{fig:SuppTrionShifts.pdf}
	\end{figure}
	\begin{figure*}
		\centering
		\begin{subfigure}[b]{0.475\textwidth}   
			\centering 
			\includegraphics[width=\textwidth]{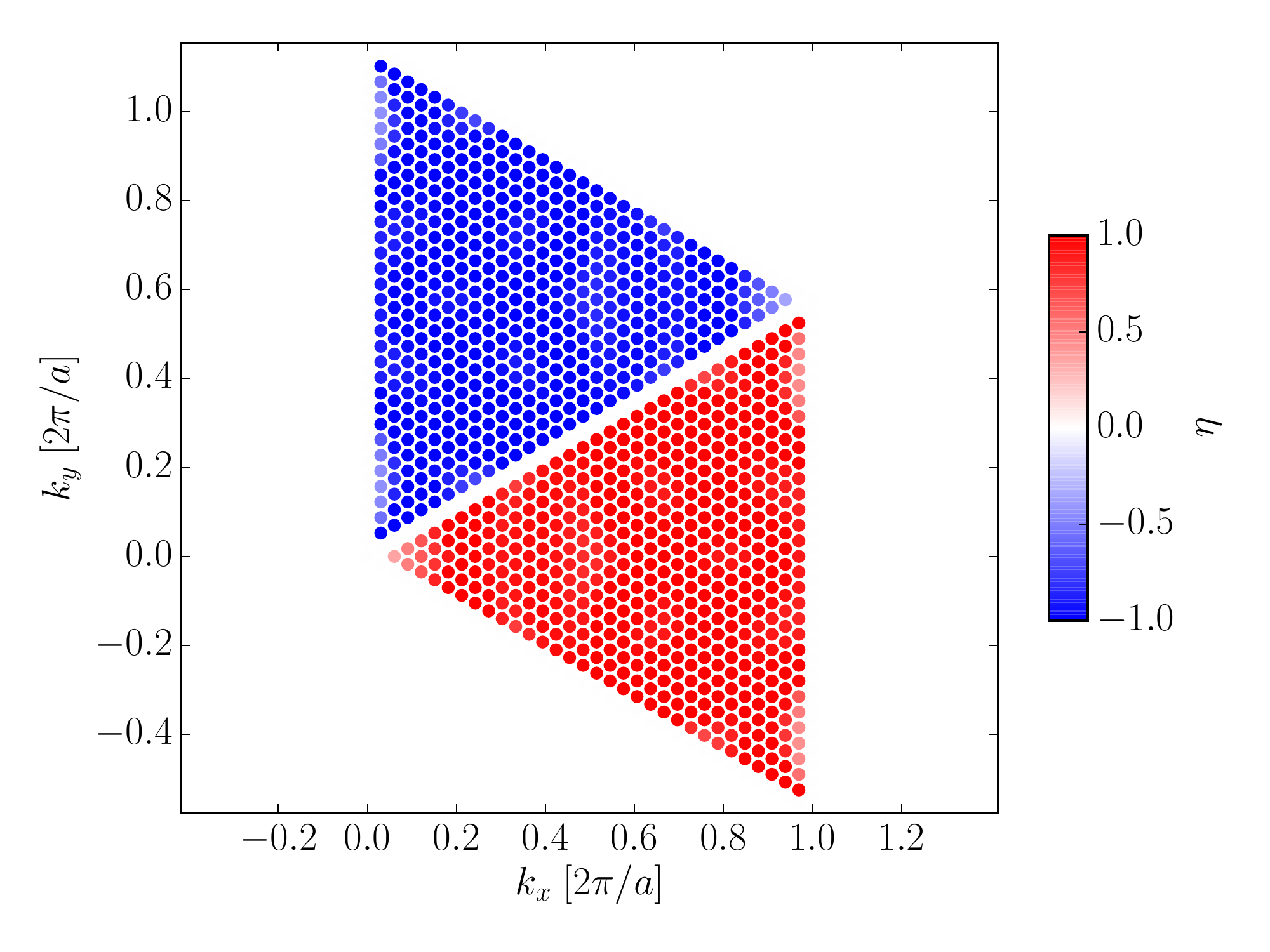}
			\caption[]%
			{{\small $\eta(\bold{k})$}}    
			\label{fig:circ_pol}
		\end{subfigure}
		\hfill
		\begin{subfigure}[b]{0.475\textwidth}
			\centering
			\includegraphics[width=\textwidth]{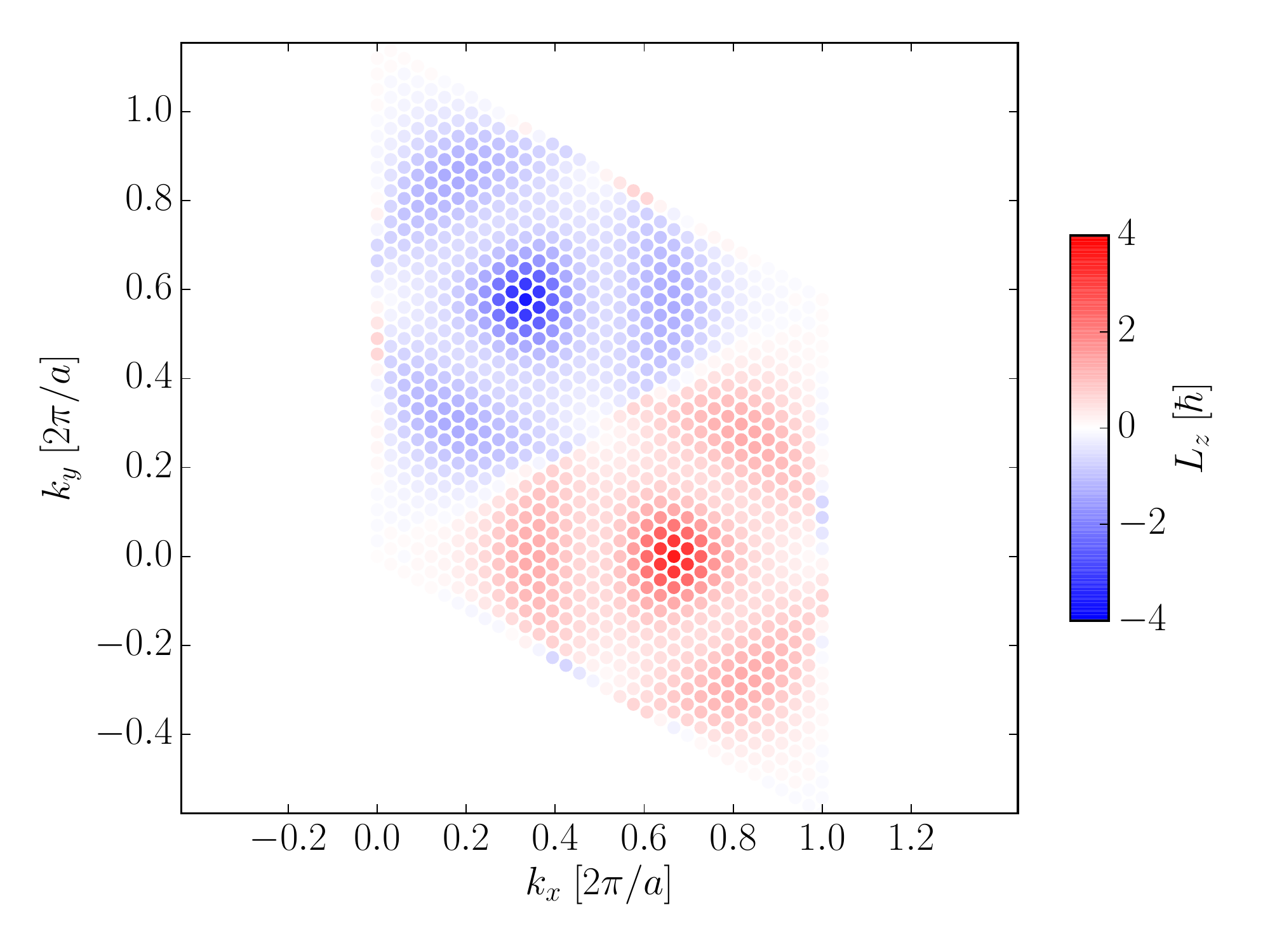}
			\caption[]%
			{{\small $L^z_\text{v}(\bold{k})$}}    
			\label{fig:orb_mom_val}
		\end{subfigure}
		\vskip\baselineskip
		\begin{subfigure}[b]{0.475\textwidth}  
			\centering 
			\includegraphics[width=\textwidth]{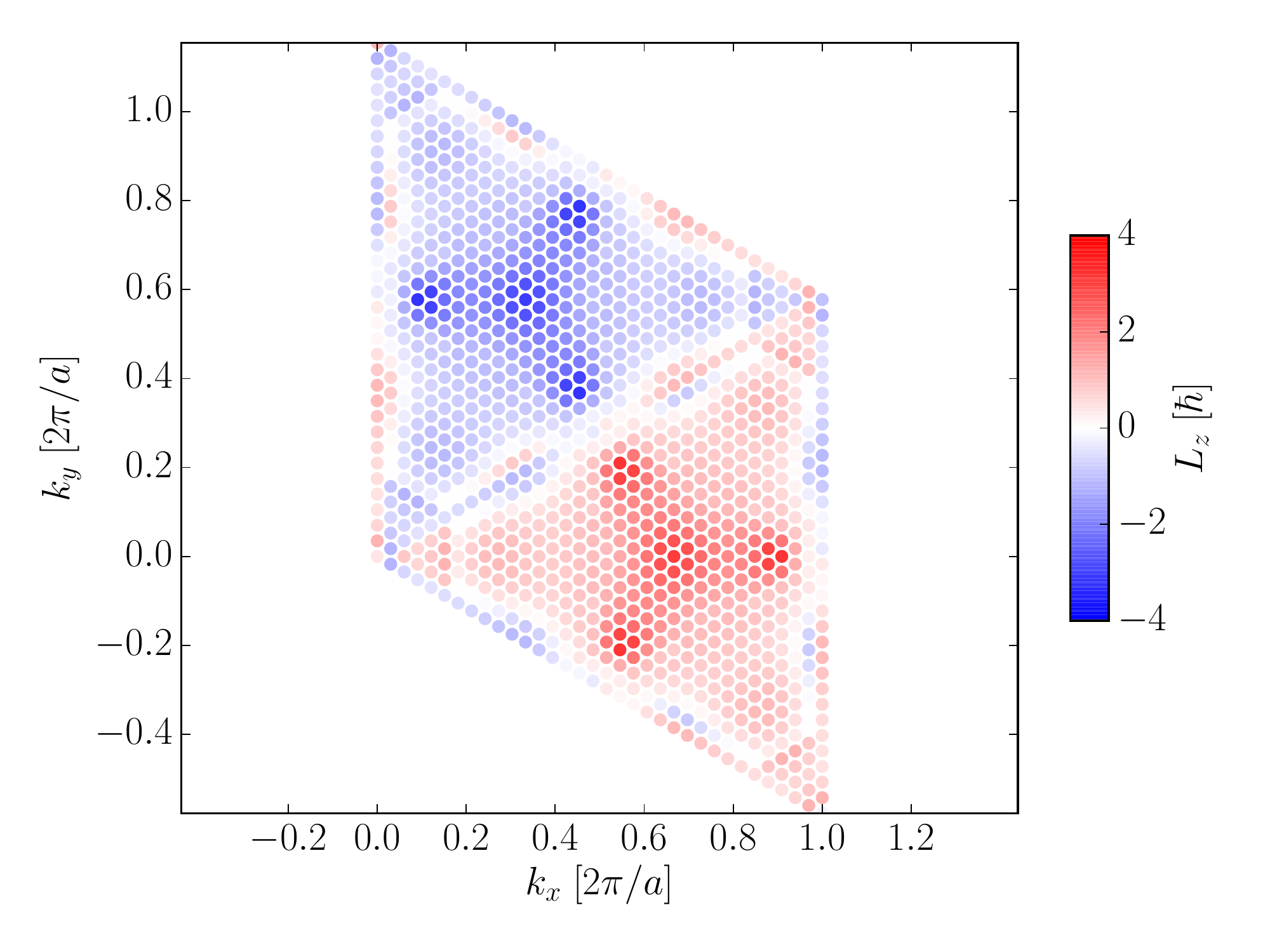}
			\caption[]%
			{{\small $L^z_\text{c}(\bold{k})$}}    
			\label{fig:orb_mom_cond}
		\end{subfigure}
		\quad
		\begin{subfigure}[b]{0.475\textwidth}   
			\centering 
			\includegraphics[width=\textwidth]{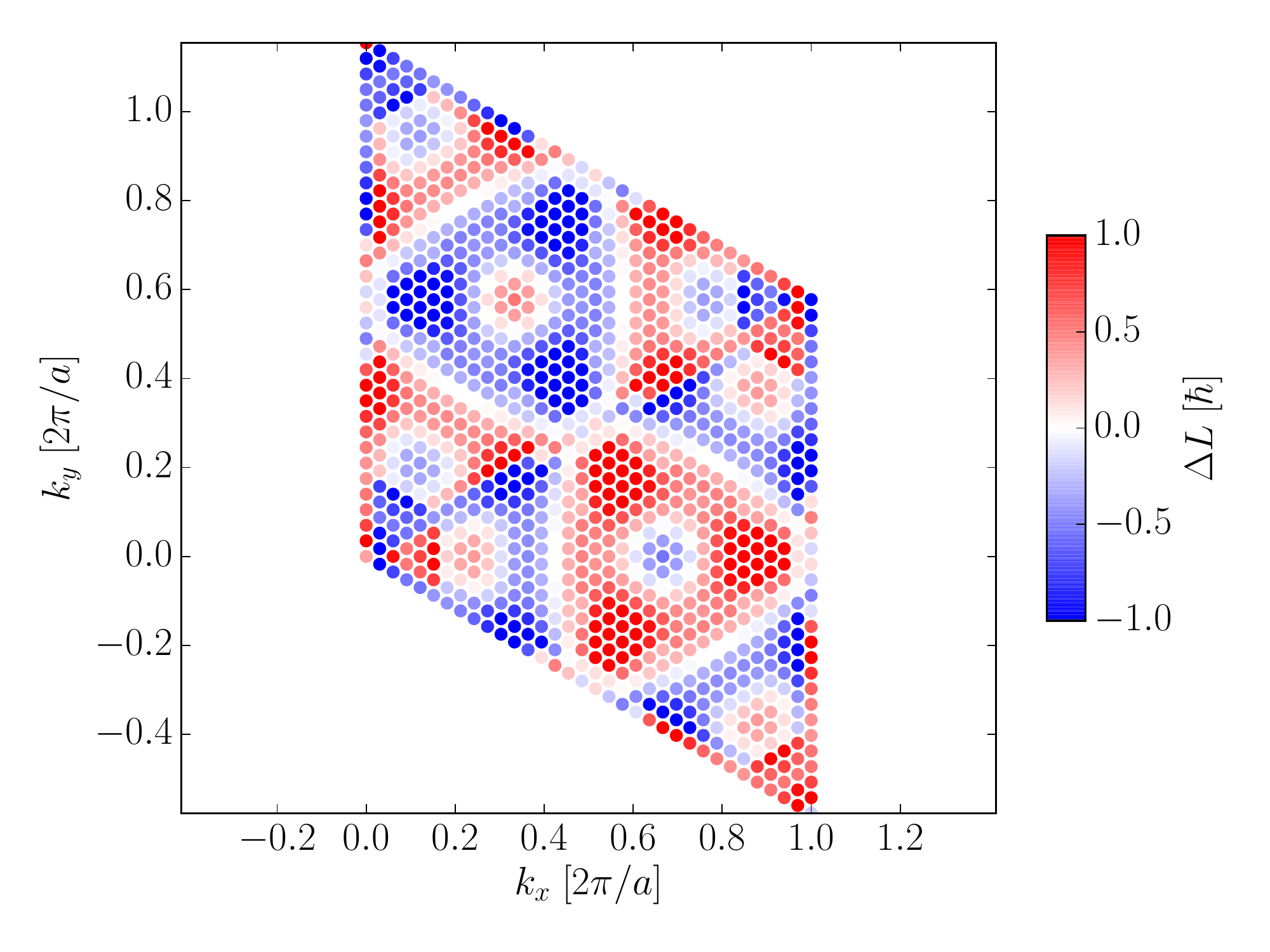}
			\caption[]%
			{{\small $\Delta L = L^z_\text{c}(\bold{k}) - L^z_\text{v}(\bold{k})$}}    
			\label{fig:exc_orb_mom}
		\end{subfigure}
		\caption{\footnotesize\textbf{Ab-initio calculated quantities in the first Brillouin zone of WS$_2$.} (a) Degree of circular polarization $\eta(\bold{k})$ (no spin-orbit coupling), (b) angular momentum of the higher-energy valence subband $L^z_\text{v}(\bold{k})$, (c) angular momentum of higher-energy conduction subband $L^z_\text{c}(\bold{k})$, and (d) difference between conduction and valence band angular momentum $\Delta L = L^z_\text{c}(\bold{k}) - L^z_\text{v}(\bold{k})$. The +K point resides at ($\frac{4\pi}{3a}$, 0).} 
		\label{fig:mean and std of nets}
	\end{figure*}
	\clearpage
	\newpage

	\section{Supplementary Notes}
	\subsection{Supplementary Note 1: Power dependence of the PL spectra in high magnetic fields}
	In order to identify spectral features that could be related to biexciton emission, we performed PL measurements at two different excitation powers in a high magnetic field of 30\,T. In Supplementary Fig. \ref{fig:Supp_PowerComparison} normalized PL spectra for excitation densities of 0.4\,kW/cm$^2$ (red curve) and 64\,kW/cm$^2$ (black curve) taken at a temperature of 4\,K and a field of 30\,T are directly compared. The PL peak positions differ slightly for the two measurements due to small variations in the measurement spot and the optically increased quasiparticle density for the high excitation density. In relation to the exciton and trion peaks, the L$_1$/XX peak clearly gains in spectral weight for the high excitation power, indicating a superlinear power-dependence of the corresponding PL intensity, which is reminiscent of biexciton emission \cite{Kim1994, You2015a}. We note that even at the relatively low cw-excitation densities of 64\,kW/cm$^2$, as used in this experiment, the biexcitonic behavior of the L$_1$/XX peak in the WS$_2$ spectrum is observable, in line with recent reports \cite{Plechinger2015a}.

	In contrast, we see no transfer of spectral weight to the X$_\text{U}$ upon an increase of the excitation power. Instead of a clearly pronounced, sharp X$_\text{U}$ peak at the low excitation density, we observe only a broad spectral feature comprised of trion and X$_\text{U}$ for the high excitation density. A biexcitonic origin of the X$_\text{U}$ peak would imply a superlinear power dependence. Therefore, it should dominate the high excitation density spectrum way more than in the presented data. The X$_\text{U}$ behaves rather like the exciton and trion features. Hence, a biexcitonic origin of the X$_\text{U}$ peak is unlikely.

	\subsection{Supplementary Note 2: Magnetic-field-induced energy shifts in the lower-energy conduction band}

	The lower-energy conduction bands in Tungsten-based TMDCs experience an energy shift due to two different magnetic moments: first, the valley magnetic moment, associated with the self-rotation of the Bloch wave-packets, and second, the pure spin magnetic moment of the electrons. The overall shift is then given by $\Delta E_c=(\tau\alpha_c + s_z)\mu_B B$, with $\tau$ being the valley index, which is +1 for +K and -1 for -K, $s_z$ the spin orientation ($+1$ for spin up and $-1$ for spin down) and $\alpha_c=L^z_c/\mu_B$, where $L^z_c$ represents the valley magnetic moment of the conduction band and $\mu_B$ the Bohr magneton. The lower-energy states of the spin-split conduction bands at the +K and -K points are the ones where the excess electron is located \cite{Yu2014, Plechinger2016a}. Their splitting in a magnetic field amounts then to $2(\alpha_c-1)\mu_B B = g'\mu_B B$. In order to describe the splitting $\Sigma$ between singlet and triplet trion state in a magnetic field, we have to take into account the Zeeman shift of the excitons and the shift of the excess electrons. Therefore, we have \cite{Aivazian2014}:
	\begin{equation} \Sigma= \delta - \tau g'  \mu_B B , \end{equation}
	with $\delta$ being the zero-field singlet-triplet splitting and the valley index $\tau$ being assigned to the valley where the initial electron-hole-pair is located.
	
	A summary of all magnetic-field-induced band-shifts in a single-particle-picture and the electron configurations for the different trion species is given in Supplementary Fig. \ref{fig:SuppTrionShifts.pdf}(a) and (b). The changes of the singlet-triplet splitting $\Sigma$ in a magnetic field are illustrated in Supplementary Fig. \ref{fig:SuppTrionShifts.pdf}(c).

	\subsection{Supplementary Note 3: Valley and orbital Zeeman angular momentum from ab-initio calculations}
	
	In previous studies a three-band tight-binding model was used to estimate orbital effects in WSe$_2$ \cite{Srivastava2014}. In order to describe contributions to the valley Zeeman effect in WS$_2$ we carry out ab-initio calculations to obtain the momentum-resolved self-rotating angular momentum expectation values \cite{Chang1996}
	
	\begin{align*}
	L^z_n(\bold{k}) &= \frac{m}{i \hbar} \sum_{l\neq n} \frac{\bra{n,\bold{k}}\frac{\partial H}{\partial k_x} \ket{l, \bold{k}} \bra{l,\bold{k}}\frac{\partial H}{\partial k_y} \ket{n, \bold{k}}}{E_n - E_l} - c.c.\\
	&\approx \frac{\hbar}{i m} \sum_{l\neq n} \frac{\bra{n,\bold{k}}\bold{p}_x \ket{l, \bold{k}} \bra{l,\bold{k}}\bold{p}_y \ket{n, \bold{k}}}{E_n - E_l} - c.c.\,.
	\end{align*}
	
	We approximate the velocity operator by momentum matrix elements obtained by the plane-wave DFT code \textsc{quantum espresso} \cite{Giannozzi2009}. In our calculations we use a lattice constant of 3.191~\AA\, and obtain a band gap of 1.89~eV using the PBE exchange-correlation functional \cite{Perdew1996}.
	
	Additionally we calculate the degree of optical polarization \cite{Cao2012}
	\begin{align*}
	\eta(\bold{k}) &= \frac{|\mathcal{P}_+^\text{cv}(\bold{k})|^2 - |\mathcal{P}_-^\text{cv}(\bold{k})|^2}{|\mathcal{P}_+^\text{cv}(\bold{k})|^2 + |\mathcal{P}_-^\text{cv}(\bold{k})|^2},\\
	\mathcal{P}_\pm^\text{cv} &= \frac{1}{\sqrt{2}} (P_x^\text{cv} \pm P_y^\text{cv}),\\
	\bold{P}^\text{cv}(\vec{k}) &= \bra{c, \bold{k}} \bold{p} \ket{v, \bold{k}},
	\end{align*}
	
	which serves as a cross-check calculation for the momentum matrix elements. In Supplementary Fig. \ref{fig:circ_pol} we show the calculated degree of optical polarization $\eta$, which agrees very well with results of Ref. \cite{Cao2012}, being positive around the +K and negative around the $-$K point.
	
	We calculated the self-rotating angular momentum expectation values for higher-energy spin-split subbands $L^z_n(\bold{k})$, where $n$ labels the valence (v) and conduction band (c), which are shown in Supplementary Figs. \ref{fig:orb_mom_val} and \ref{fig:orb_mom_cond}, respectively. In our calculations we took into account 80 bands (in total 18 valence and 62 conduction bands), including spin-orbit coupling. We obtain values of $L^z_\text{v}(+\mathrm{K}) = 3.580\,\hbar$ and $L^z_\text{c}(+\mathrm{K}) = 3.038\,\hbar$, yielding an effective contribution of $\Delta L = L^z_\text{c}(\mathrm{+K}) - L^z_\text{v}(\mathrm{+K}) = -0.543\,\hbar$. Additionally to the self-rotating angular momentum, we evaluate expectation values of projections on spherical harmonics centered at the atom positions to get the intracellular contributions to the angular momentum. In tight-binding theory one finds that the states in question are composed out of pure $d$ orbitals yielding $m_z = \pm 2$ and $m_z=0$ for the valence and conduction band, respectively \cite{Liu2013}. From the ab-initio point of view one expects also contributions from $p$ orbitals of the S atoms. We find $m_z = \pm 1.688$ and $m_z = \mp 0.142$ for the valence and conduction bands, respectively. This reduces the intracellular angular momentum contribution from $2\,\hbar$ to $1.830\,\hbar$.
	
	In total we obtain a g factor of $-2(1.830-\Delta L/\hbar)=-4.746$. We believe our calculation can give a semi-quantitative explanation of the enhancement of the g factor. To obtain fully quantitative results, one should take into account the final spread of the exciton wave function in $k$ space as well as incorporate spin-orbit coupling contributions to the velocity operator.

	\subsection{Supplementary Note 4: Model for the theoretical description of a 2D exciton}
	
	Excitons in transition metal dichalcogenide monolayers cannot be described by a simple hydrogen model as the anisotropic dielectric environment of the flake has to be accounted for. The binding energy and the wavefunctions are developped in the framework of an envelope approximation, which describes the relative motion of electron and hole in a Wannier exciton \cite{Poellmann2015, Chernikov2014}. After a separation of the variables, the resulting Schr\"odinger equation for the radial part of the envelope function $\psi(r)$ reads:
	\begin{equation} \frac{\partial^2}{\partial r^2} \psi(r) + \frac{1}{r}\frac{\partial}{\partial r} \psi(r) + \left[\frac{2\mu}{\hbar} \left(E-V_{eh}(r)\right) - \frac{l^2}{r^2} \right] \psi(r) = 0 ,\end{equation}
	with the energy E, the orbital quantum number $l$, and the electron-hole-potential $V_{eh}(r)$. For s-states with $l=0$ and with the ansatz $\psi(r)=r^{-1/2}u(r)$, we get a 1D Schr\"odinger equation:
	\begin{equation} \left[-\frac{\hbar^2}{2\mu}\frac{\partial^2}{\partial r^2} + V_{eff}(r) - E \right]u(r) = 0. \label{eq:SuppSchroedinger}\end{equation}
	Here, the effective potential $V_{eff}$ reads as follows:
	\begin{equation} V_{eff}(r)=V_{eh}(r)-\frac{\hbar^2}{8\mu r^2}. \end{equation}
	In 2D semiconductors, $V_{eh}$ differs considerably from the hydrogenic $1/r$ potential. Instead, a potential proposed by Keldysh accurately accounts for the non-uniform dielectric environment of the monolayer \cite{Keldysh1976}:
	\begin{equation} V_{eh}^{2D}(r)=-\frac{\pi e^2}{2r_0}\left[ H_0 \left( \frac{r}{r_0} \right) - Y_0 \left( \frac{r}{r_0} \right) \right],  \end{equation}
	with the Struve and Neumann function of order 0 and the screening length $r_0$. 
	
	As introduced by Chernikov et al. \cite{Chernikov2014}, the crossover between the long-range $1/r$ and the short-range $\log (r)$ Coulomb interaction is quantified by $r_0$. Chernikov et al. have shown that with an adjusted screening length of 75\,\AA$\:$(in contrast to 38\,\AA$\:$for the bare flake), the exciton model describes very accurately the experimental exciton spectra of monolayer WS$_2$ on a SiO$_2$ substrate. Taking a reduced mass of $\mu=0.15\,m_e$ \cite{Kormanyos2015}, the numerically calculated solution of the Schr\"odinger equation (Supplementary Equation \ref{eq:SuppSchroedinger}) yields for the $1s$ ground state a binding energy of 312\,meV, a Bohr radius of 16\,\AA$\:$ and a root-mean-square exciton radius of 25\,\AA.

	\bibliographystyle{achemso}
	\bibliography{library}

\end{document}